\documentclass[letterpaper,twocolumn,english,aps,pra,floatfix]{revtex4}
\usepackage{color}
\usepackage{verbatim}
\usepackage{amsmath}
\usepackage{amssymb}
\usepackage{graphicx}

\newcommand{\ucite}[1]{\textsuperscript{\cite{#1}}}
\makeatletter

\usepackage{dcolumn}% Align table columns on decimal point
\usepackage{bm}% bold math
\usepackage[colorlinks,citecolor=blue, linkcolor=blue,hyperindex,CJKbookmarks,dvipdfm]{hyperref}
\begin{document}

\vspace{2.5cm}
\title{Distribution of quantum Fisher information in asymmetric cloning machines}
\author{Xing Xiao$^{1,2}$}
\altaffiliation{xiaoxing1121@gmail.com}
\author{Yao Yao$^{2}$}
\author{Lei-Ming Zhou$^{2,3}$}
\author{Xiaoguang Wang$^{4}$}
\altaffiliation{xgwang@zimp.zju.edu.cn}
\affiliation{${1}$ College of Physics and Electronic Information, Gannan Normal University, Ganzhou 341000, China\\
$2$ Beijing Computational Science Research Center, Beijing 100084, China \\
$3$ Key Laboratory of Quantum Information, University of Science and Technology of China, Hefei 230026, China\\
$4$ Zhejiang Institute of Modern Physics, Department of Physics, Zhejiang University, Hangzhou 310027, China}

\begin{abstract}
An unknown quantum state cannot be copied on demand and broadcast freely due to
the famous no-cloning theorem. Approximate cloning schemes have
been proposed to achieve the optimal cloning characterized by the maximal fidelity
between the original and its copies. Here, from the perspective of quantum Fisher information (QFI),
we investigate the distribution of QFI in asymmetric cloning machines which produce two nonidentical copies.
As one might expect, improving the QFI of one copy results in decreasing the QFI of
the other copy, roughly the same as that of fidelity. It is perhaps also unsurprising that
asymmetric phase-covariant cloning machine outperforms universal cloning machine
in distributing QFI since a priori information of the input state has been utilized.
However, interesting results appear when we compare the distributabilities of fidelity (which quantifies the full information of quantum states), and
QFI (which only captures the information of relevant parameters) in asymmetric cloning machines.
In contrast to the results of fidelity, where the distributability of
symmetric cloning is always optimal for any $d$-dimensional cloning, we find that asymmetric cloning
performs always better than symmetric cloning on the distribution of QFI for $d\leq18$, but this conclusion becomes invalid when $d>18$.

%The better or worse of asymmetric cloning depends on the parameters of asymmetric cloning.

%capability (measured by the sum of two copies' fidelities and QFIs)
%of distributing information in asymmetric cloning machines.

%
%However, unlike the results of fidelity,
%which the sum of two replicates' fidelities in asymmetric cloning is always smaller than the symmetric cloning for $d$-dimensional cloning,
%the results of QFI is more subtle. For $d\leq18$, the sum of QFIs in asymmetric cloning is always larger than that in symmetric cloning case,
%which means that asymmetric cloning always performs better than symmetric cloning in distributing QFI under this condition,
%but this conclusion  becomes invalid when $d>18$. It may be larger or smaller than the case of
%symmetric cloning, which depends on the parameter of asymmetric cloning.

\end{abstract}
%\pacs{03.67.-a,03.67.Ta,03.67.Hk}
%\keywords{}
\maketitle

Classical information can be replicated perfectly and broadcast
without fundamental limitations. However, information encoded in
quantum states is subject to several intrinsic restrictions of
quantum mechanics, such as Heisenberg's uncertainty
relations\ucite{heisenberg27} and quantum no-cloning
theorem\ucite{wootters82}. The no-cloning theorem tells us that an
unknown quantum state cannot be perfectly replicated because of the
linearity of the time evolution in quantum physics, which is the
essential prerequisite for the absolute security of quantum
cryptography\ucite{gisin02a}. Nevertheless, it is still possible to
clone a quantum state approximately, or instead, clone it perfectly
with certain probability\ucite{scarani05,fan14}. Therefore, various
types of quantum cloning machines have been designed for different
quantum information tasks, including universal quantum cloning machine (UQCM)\ucite{buzek96,gisin97},
state-dependent cloning machines\ucite{bruss98} and phase-covariant quantum cloning machine
(PQCM)\ucite{niu99,bruss00,fan03,du05}.

So far, the optimality of the approximate cloning machine is judged
generally by whether the obtained fidelity between the cloning output
state and the ideal state achieves its optimal bound. Although the fidelity may have
qualified the complete information of the quantum states, in most
scenarios, only the information of certain parameters which are
physically encoded in quantum states is our practical concern. For
example, the relative phase estimation is an extremely important
issue in the field of quantum metrology\ucite{giov06,giov11}. Thus
it is not necessary to gain complete information of the whole
quantum states themselves, but rather the relevant parameter
information. QFI is a natural candidate to quantify the physical
information about the involved parameters\ucite{helstrom76}. In ref.
\cite{lu13}, the authors pointed that the QFI of relevant parameter
encoded in quantum states also cannot be cloned perfectly, while it
might be broadcast even in some non-commuting quantum states.
Furthermore, from the perspective of QFI, Song \emph{et al.} showed
that Wootters-Zurek cloning performs better than universal cloning
for the symmetric cloning cases\ucite{song13}. In our recent work,
the multiple phase estimation problem was investigated
in the framework of symmetric quantum cloning machines\ucite{yao14}.

On the other hand, we note that quantum cloning machines not only provide a
good platform for investigating distribution of quantum information,
but also have been proved to be very efficient eavesdropping attacks
on the quantum key distribution (QKD)
protocols\ucite{fuchs97,bruss02,cerf02,xiong12}. In this context,
asymmetric quantum cloning machines would be of particular interest
since the eavesdropper can adjust the trade-off between the
information gained from a quantum communication channel and the
error rate of information transmitted to the authorized receiver.
Motivated by these considerations, we investigate the problem of
distributing QFI in asymmetric quantum cloning machines for any
dimension. We focus on the following four questions: (i) Is it
possible to improve the QFI of one copy by decreasing that of the
other copy? If YES, what's the trade-off relation between them? (ii)
Does asymmetric PQCM always perform better than asymmetric UQCM on
the capability of distributing QFI? (iii) Does asymmetric cloning
always outperform symmetric cloning in distributing QFI for any
dimensionality? (iv) What's the difference between fidelity and QFI on
the characterization of distributability in asymmetric cloning? Except
for the fourth question need to be clarified in detail, we can
briefly answer the first two questions in the affirmative but the
third in
the negative. %That is because asymmetric cloning is not always better symmetric when the dimension is larger than 18.
Our results shed an alternative light on quantum cloning and may be exploited for quantum phase estimation.

%
%
%Within this framework we discuss the problem of distributing QFI and compare
%with the case of symmetric quantum cloning.
%We show that similar to the behavior of fidelity, QFI also can be
%improved in one copy by paying the price of the decrease of QFI in the other copy. However, more interesting results
%appear when we consider the sum of fidelities and QFIs of two copies. We find that the sum of two copies' fidelities in
%asymmetric cloning is always smaller than that in the symmetric cloning case for both qubit and qudit ($d>2$) cloning.
%In contrast to the fidelity, for the qubit cloning, the sum of QFIs always attains its minimum value only when the
%asymmetric cloning reduces to the symmetric case. While the results become subtle for qudit cloning and a critical
%point would occur when $d=18$, which means the above conclusion only holds for $d\leq18$ and fails for larger $d$.

%counterintuitive

%
%\begin{figure}
%\label{fig1}
%  \centering
%  \includegraphics[width=3.2in,height=1.4in]{fig0.eps}\\
%  \caption{Schematics of symmetric quantum cloning and asymmetric quantum cloning}
%\end{figure}

\section*{Results}
\textbf{Quantum Fisher information.} We start with a brief
introduction of QFI and give a useful form of QFI for a special kind
of mixed qudit states, which usually represents the output states of
qudit cloning. Recall that QFI of parameter $\theta$ encoded in
$d$-dimensional quantum state $\rho_{\theta}$ is generally defined
as\ucite{helstrom76,holevo82,braun94}
\begin{equation}
\label{e1}
\mathcal {F}_{\theta}=\textrm{Tr}(\rho_{\theta}\mathcal {L}_{\theta}^{2}),
\end{equation}
where $\mathcal {L}_{\theta}$ is the so-called symmetric logarithmic
derivative, which is defined by
$\partial_{\theta}\rho_{\theta}=(\mathcal
{L}_{\theta}\rho_{\theta}+\rho_{\theta}\mathcal {L}_{\theta})/2$
with $\partial_{\theta}=\partial/\partial\theta$. By diagonalizing
the matrix as
$\rho_{\theta}=\Sigma_{n}\lambda_{n}|\psi_{n}\rangle\langle\psi_{n}|$,
one can rewritten the QFI as\ucite{knysh11,liu13}
\begin{equation}
\label{e2}
\mathcal {F}_{\theta}=\sum_{n}\frac{(\partial_{\theta}\lambda_{n})^2}{\lambda_{n}}+\sum_{n}\lambda_{n}\mathcal{F}_{\theta,n}
-\sum_{n\neq m}\frac{8\lambda_{n}\lambda_{m}}{\lambda_n+\lambda_m}|\langle\psi_{n}|\partial_{\theta}\psi_{m}\rangle|^2,
\end{equation}
where $\mathcal {F}_{\theta,n}$ is the QFI for pure state $|\psi_{n}\rangle$ with the form
\begin{equation}
\mathcal {F}_{\theta,n}=4[\langle\partial_{\theta}\psi_{n}|\partial_{\theta}\psi_{n}\rangle-|\langle\psi_{n}|\partial_{\theta}\psi_{n}\rangle|^2].
\end{equation}
Note that Eq. (\ref{e2}) suggests the QFI of a non-full-rank state
is only determined by the subset of $\{|\psi_{i}\rangle\}$ with
nonzero eigenvalues. Physically, the QFI can be divided into three
parts\ucite{zhang13,liu13}. The first term is just the classical
Fisher information determined by the probability distribution; The
second term is a weighted average over the QFI for all the nonzero
eigenstates; The last term stemming from the mixture of pure states
reduces the QFI and hence the estimation precision below the
pure-state case. Though the Eq. (\ref{e2}) is powerful, there is no
explicit expression for an arbitrary $d$-dimensional mixed state.
However, it is worth noting that the output reduced states of UQCM
and PQCM all have a form as
\begin{equation}
\label{e3}
\rho_{out}=\eta|\psi\rangle_{in}\langle\psi|+\frac{1-\eta}{d}\textrm{I}_{d},
\end{equation}
which is completely characterized by a parameter independent
shrinking factor $\eta$ and the dimensionality $d$. Here, $|\psi\rangle_{in}$ and $\textrm{I}_{d}$
are the input state and $d$-dimensional identity matrix,
respectively. Although $\rho_{out}$ has such a simple form, an analytical
expression of QFI is still difficult to achieve. Fortunately,
if we restrict our discussions to the special case of
input states in the form
\begin{equation}
\label{e4}
|\psi\rangle_{in}=\frac{1}{\sqrt{d}}\sum_{k}e^{i\theta_{k}}|k\rangle,
\end{equation}
which are covariant with respect to rotations of the phases, a general form of QFI for any parameter $\theta_{k}$
could be given by (see methods)
\begin{equation}
\label{e5}
\mathcal {F}_{\theta_{k}}=\frac{4(d-1)\eta^2}{2d+d(d-2)\eta}.
\end{equation}

The above equation is a key mathematical tool for our analysis of this paper.
Although this expression only holds for the combination of Eqs. (\ref{e3}) and (\ref{e4}),
it is powerful since the scaling form of $\rho_{out}$ is usually satisfied in quantum cloning machines
or in the case of a pure state under white noises. On the other hand, the equatorial states are widely employed
in the physical implementations of quantum communication
ideas (such as BB84 protocol\ucite{bb84}) as well as in the demonstration of fundamental
questions in quantum information processing.
After a simple calculation,
we find that $\mathcal {F}_{\theta_{k}}$ is a monotonically
increasing function of the shrinking factor $\eta$. This is to be
expected because the larger $\eta$ indicates more
information the reduced output state $\rho_{out}$ contains about
the relevant parameter.
%
%Although this expression only holds for the combination of Eqs. (\ref{e3}) and (\ref{e4}),
%it will be a useful tool for our analysis of QFI distribution, since
%these two conditions are usually satisfied in quantum cloning machines.

\textbf{Distributability.} Before moving to the discussion of distribution,
a proper measure that quantify the distributability of cloning machines should be well defined.
Note that one can define the distributability in different ways
which depends on what is distributed in the procedure. For example,
it can be defined from the perspectives of fidelity which quantifies
the total information of the state, and QFI which quantifies the information
of particular parameters in the state.

It is well known that, unlike the symmetric cloning in which all the
copies are the same, the outputs of asymmetric cloning are
nonidentical. Hence, the optimality of asymmetric cloning can be
judged by maximizing the sum of all copies' fidelities, as discussed
in refs. \cite{cerf00,ghiu03}. Intuitively, when we consider the
distribution of quantum states, the distributability of asymmetric
cloning can be defined as
\begin{equation}
\label{e06}
F=\sum_{i}F_{i},
\end{equation}
where $F_{i}$ is the fidelity between the original and the $i$th copy.
Namely, the larger $F$ indicates the better capability of distribution on the quantum states.

In a seminal work\ucite{braun94}, the authors pointed that both
fidelity and QFI are highly related to the distinguishability of the
states, which is measured by Bures distance\ucite{bures69}.
Therefore, from the perspective of QFI, the measure
\begin{equation}
\label{e07}
\mathcal{F}=\sum_{i}\mathcal{F}_{i,\theta},
\end{equation}
also qualifies the capability of distributing information of
relevant parameter encoded in the input state, where $\mathcal{F}_{i,\theta}$ denotes
the QFI of parameter $\theta$ in the $i$th copy. In the following discussions, we
will use the definitions (\ref{e06}) and (\ref{e07}) to
quantify the distributabilities of fidelity and QFI respectively in asymmetric cloning machines.

\textbf{QFI distribution for 2-dimensional cloning.} As we mentioned above,
one is particularly interested in the asymmetric cloning machines
which produce two copies with different qualities within the
framework of quantum cryptography. Two typical asymmetric cloning
machines are asymmetric UQCM\ucite{cerf00}, which clones all input
states equally well, and asymmetric PQCM\ucite{niu99}, which works
equally well only for equatorial input states with the form of Eq.
(\ref{e4}). We first discuss the 2-dimensional cloning by obtaining the analytical
results, and then generalize it to \emph{d}-dimensional cloning by the assistance
of numerical simulations.

\textbf{Asymmetric 2-dimensional UQCM.} The $1\rightarrow 1+1$
optimal asymmetric UQCM was independently proposed by Niu and
Griffiths\ucite{niu98}, Bu\v{z}ek \emph{et al}\ucite{buzek98} and
Cerf\ucite{cerf00b}. Though their formalisms are slightly different,
the results are exactly the same. For the sake of convenience, we
adopt the quantum circuit approach developed by Bu\v{z}ek. The
transformation of asymmetric UQCM can be written in the following
form
\begin{align}
\label{e6}
&|\psi\rangle_{A}(a|\Phi^{+}\rangle_{BR}+b|0\rangle_{B}|+\rangle_{R})\nonumber\\
&\rightarrow a|\psi\rangle_{A}|\Phi^{+}\rangle_{BR}+b|\psi\rangle_{B}|\Phi^{+}\rangle_{AR},
\end{align}
where $|\Phi^{+}\rangle=(|00\rangle+|11\rangle)/\sqrt{2}$ and $|+\rangle=(|0\rangle+|1\rangle)/\sqrt{2}$.
The parameters $a$ and $b$ are real, and satisfy the normalization
condition $a^2+b^2+ab=1$. With the
transformation (\ref{e6}) in mind, it is now an easy exercise to verify that the two different copies
of the original state $|\psi\rangle=(|0\rangle+e^{i\theta}|1\rangle)/\sqrt{2}$ are
\begin{align}
\label{e7}
\rho_{A}&=(1-b^2)|\psi\rangle\langle\psi|+\frac{b^2}{2}\textrm{I}_{2},\\
\label{e8}
\rho_{B}&=(1-a^2)|\psi\rangle\langle\psi|+\frac{a^2}{2}\textrm{I}_{2}.
\end{align}
From the point of geometry, the asymmetric UQCM shrinks the original
Bloch vector by two different shrinking factors ($\eta_{A}=1-b^2$,
$\eta_{B}=1-a^2$) regardless of its direction. As special cases,
we can see that if $\eta_{A}=1$, then no information has been
transferred from the original system, while, if $\eta_{B}=1$, then
all of the information in system A has been transferred to system B.
In addition, if $\eta_{A}=\eta_{B}=2/3$ (i.e., $a=b=1/\sqrt{3}$),
then it reduces to the symmetric UQCM case. Obviously, the two
shrinking factors $\eta_{A}$ and $\eta_{B}$ are related, and should
satisfy the no-cloning inequality\ucite{buzek98,cerf00}
\begin{equation}
\label{e9}
\eta_{A}^2+\eta_{B}^2+(1-\eta_{A})(1-\eta_{B})\leq1,
\end{equation}
which is an ellipse in the $(\eta_{A},\eta_{B})$ space.
An optimal asymmetric UQCM is characterized by
a point $(\eta_{A}, \eta_{B})$ which lies on the boundary of this ellipse.

According to Eq. (\ref{e5}), when $d=2$, the QFI of parameter
$\theta$ is proportional to $\eta^2$. Immediately, the QFIs are
\begin{equation}
\label{e10}
\mathcal {F}_{A}=(1-b^2)^2, \mathcal{F}_{B}=(1-a^2)^2.
\end{equation}
Here and henceforth we omit the subscript $\theta_{k}$ for brevity since we restrict our discussions to
the single-parameter scenario. Remarkably, a trade-off relation exists for the two QFIs:
if one QFI is large, correspondingly another QFI will become small. Combining Eqs. (\ref{e9}) and (\ref{e10}),
the trade-off relation of QFI is expressed as
\begin{equation}
\label{e11}
\mathcal{F}_{A}+\mathcal{F}_{B}+\left(\sqrt{\mathcal{F}_{A}}-1\right)\left(\sqrt{\mathcal{F}_{B}}-1\right)\leq1.
\end{equation}
This trade-off tells us that even we only concern cloning the information of a particular parameter encoded in quantum states,
two close-to-perfect copies cannot be achieved simultaneously, imposed by quantum mechanics. An intuitive presentation of this trade-off relation
is shown in Fig.~\ref{fig2}b (dashed line).

In the following, we consider the distribution of QFI in the asymmetric UQCM.
As we defined in Eq. (\ref{e07}). the distributability of QFI is measured by
\begin{equation}
\label{e12}
\mathcal{F}^{UQCM}=\mathcal{F}_{A}+\mathcal{F}_{B}.
\end{equation}
Therefore, the larger $\mathcal{F}$ is, the more
QFI of the relevant parameter has been distributed to the two copies.
We find that the asymmetric UQCM always performs better than symmetric UQCM in distributing QFI, which means
\begin{equation}
\label{e13}
\mathcal{F}^{UQCM}(\eta_{A}\neq\eta_{B})>\mathcal{F}^{UQCM}(\eta_{A}=\eta_{B}).
\end{equation}
This can be proved by the method of Lagrange multiplier. One will
find three extreme points: $(\eta_{A}=0,\eta_{B}=1)$, $(\eta_{A}=\eta_{B}=2/3)$ and
$(\eta_{A}=1,\eta_{B}=0)$. It is easy to verify that $\mathcal{F}(\eta_{A}=\eta_{B})=8/9$ is the minimum value.

\begin{figure}
  \centering
  \includegraphics[width=0.4\textwidth]{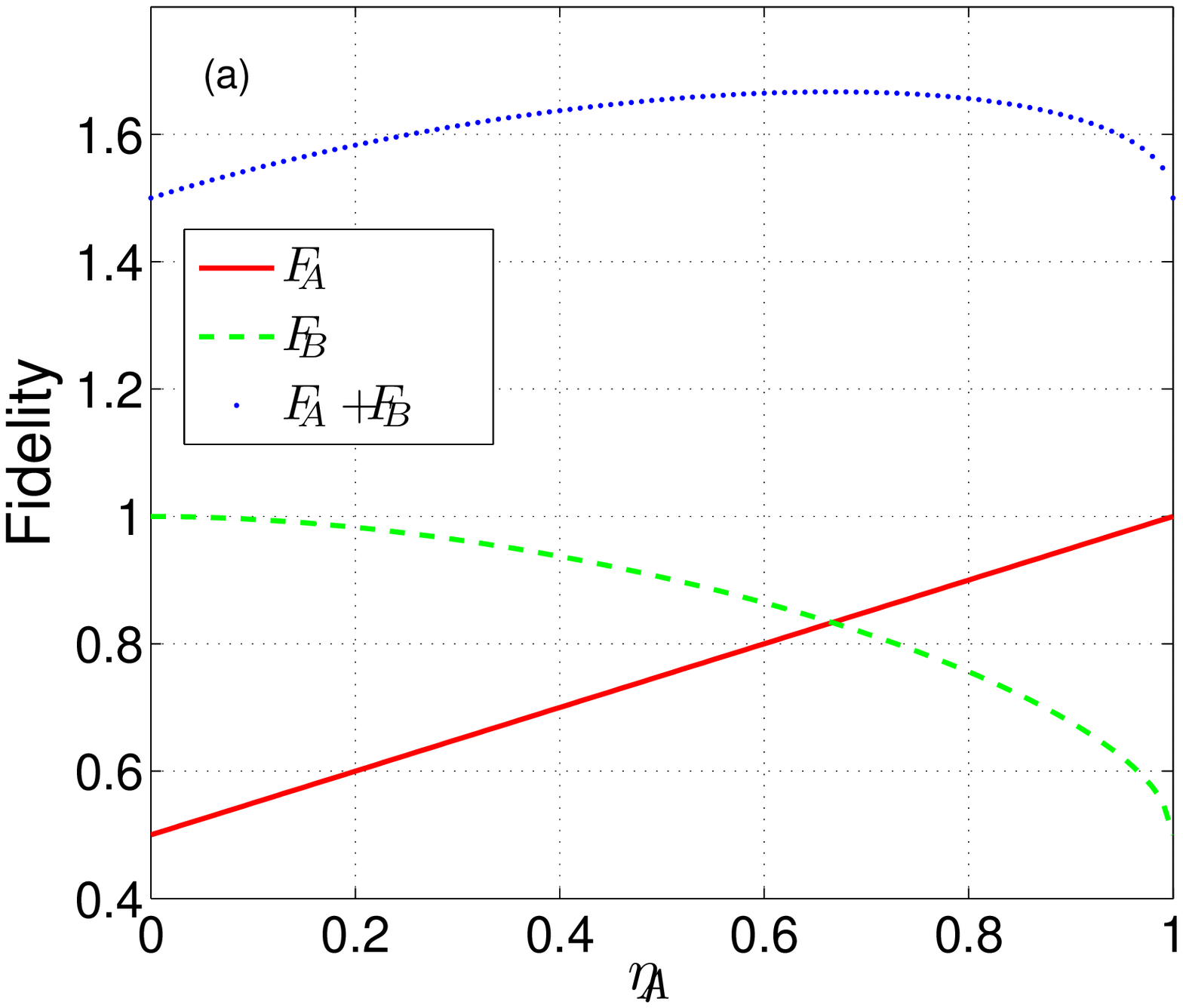}\\
  \includegraphics[width=0.4\textwidth]{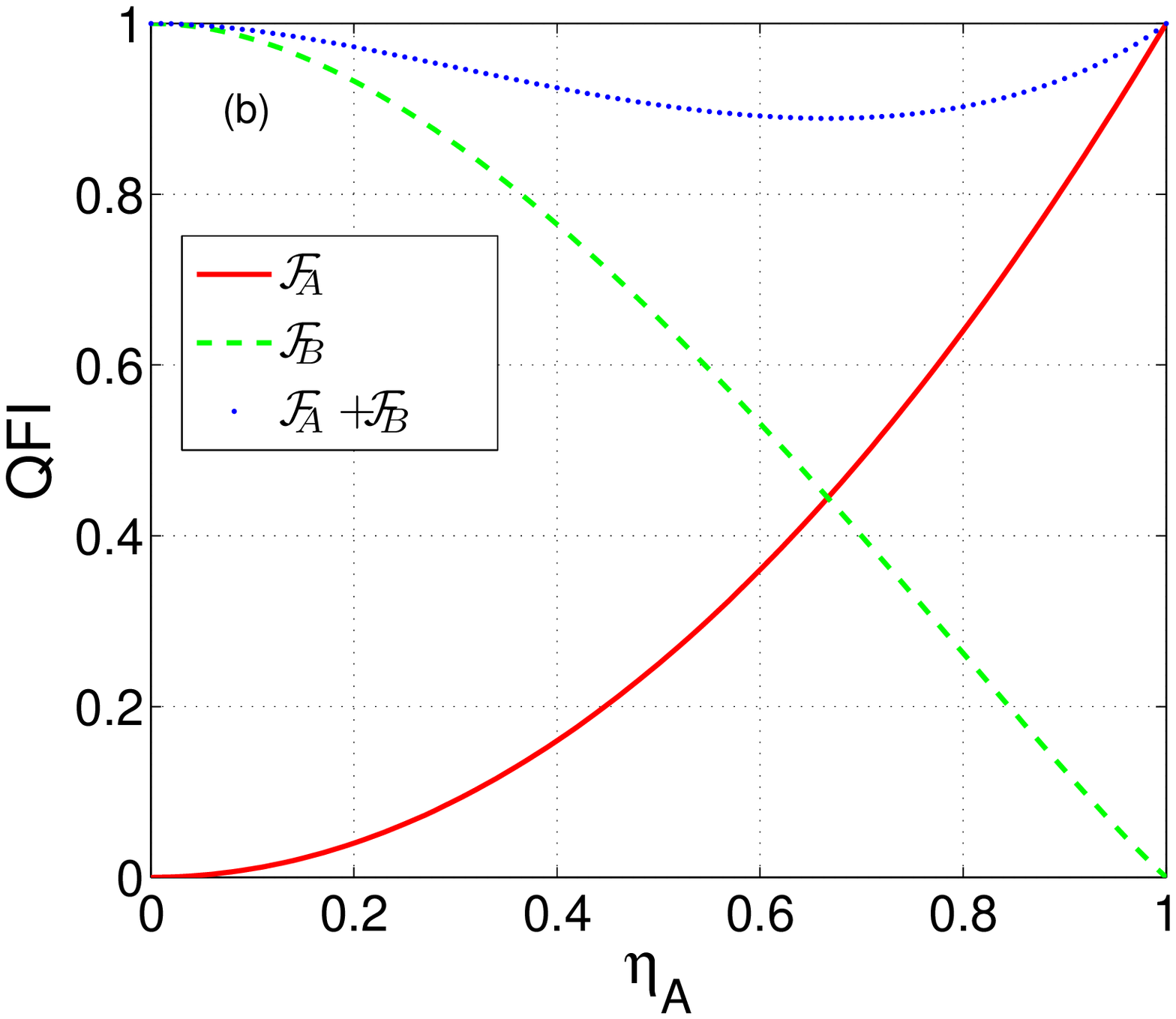}\\
  \caption{Results of asymmetric 2-dimensional UQCM: (a) Fidelities as a function of
  shrinking factor $\eta_{A}$ and (b) QFIs as a function of $\eta_{A}$.}\label{fig1}
\end{figure}

In order to show the difference of distributability between QFI and fidelity,
we also write the corresponding fidelities defined as $F_{A(B)}=\langle\psi|\rho_{A(B)}|\psi\rangle$
\begin{equation}
\label{e14}
F_{A}=1-\frac{b^2}{2}, F_{B}=1-\frac{a^2}{2}.
\end{equation}
According to Eq. (\ref{e06}), we adopt $F=F_{A}+F_{B}$ to qualify the
capability of asymmetric UQCM in distributing the entire quantum state.
Fig.~\ref{fig1} shows the results of asymmetric 2-dimensional UQCM.
It is remarkable that, from the perspective of QFI, the asymmetric
UQCM is always works better than symmetric UQCM, which is a sharp contrast to
the result of fidelity\ucite{ghiu03}, where the symmetric UQCM is always optimal.

\textbf{Asymmetric 2-dimensional PQCM.} In the context of quantum
cryptography\ucite{gisin02a}, the UQCM studied in the previous subsection might be
optimal if the detail setup of QKD protocol is not specified. But it may not be optimal for the quantum states involved
in a special QKD protocol. Practically, it is possible that we
already know a priori information of the input states. Thus, a
state-dependent cloning machine would perform better than UQCM. The
best-known example of state-dependent cloning machine is the
so-called PQCM. The symmetric PQCM was firstly proposed by Bru{\ss}
\emph{et al} for the equatorial qubit state\ucite{bruss00} and then
an asymmetric version was demonstrated by Niu and
Griffiths\ucite{niu99}. Recently, the asymmetric PQCM has been
experimentally realized using NMR\ucite{chen07} and fiber
optics\ucite{bart07}.

\begin{figure}

  \centering
  \includegraphics[width=0.4\textwidth]{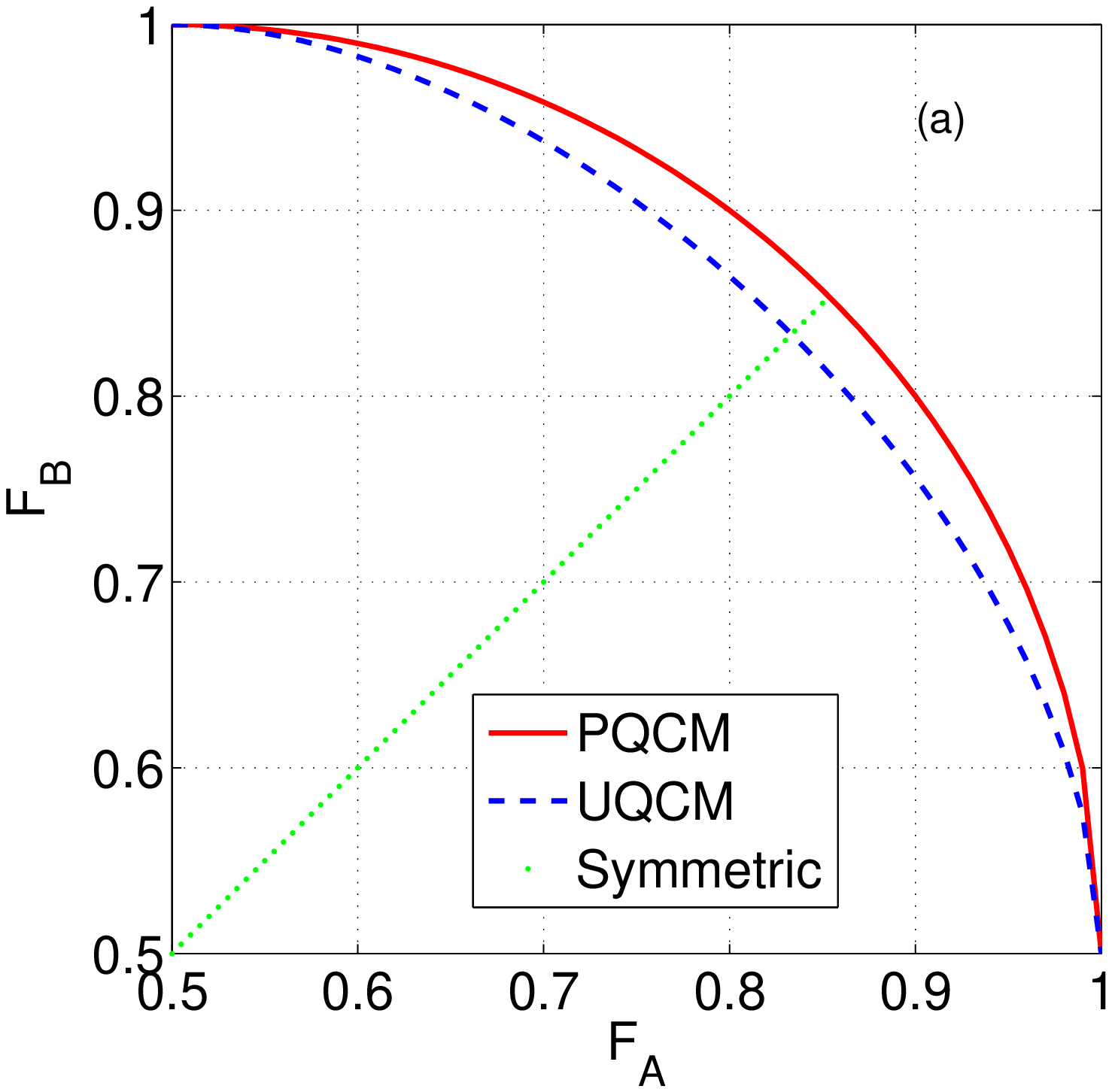}\\
  \includegraphics[width=0.4\textwidth]{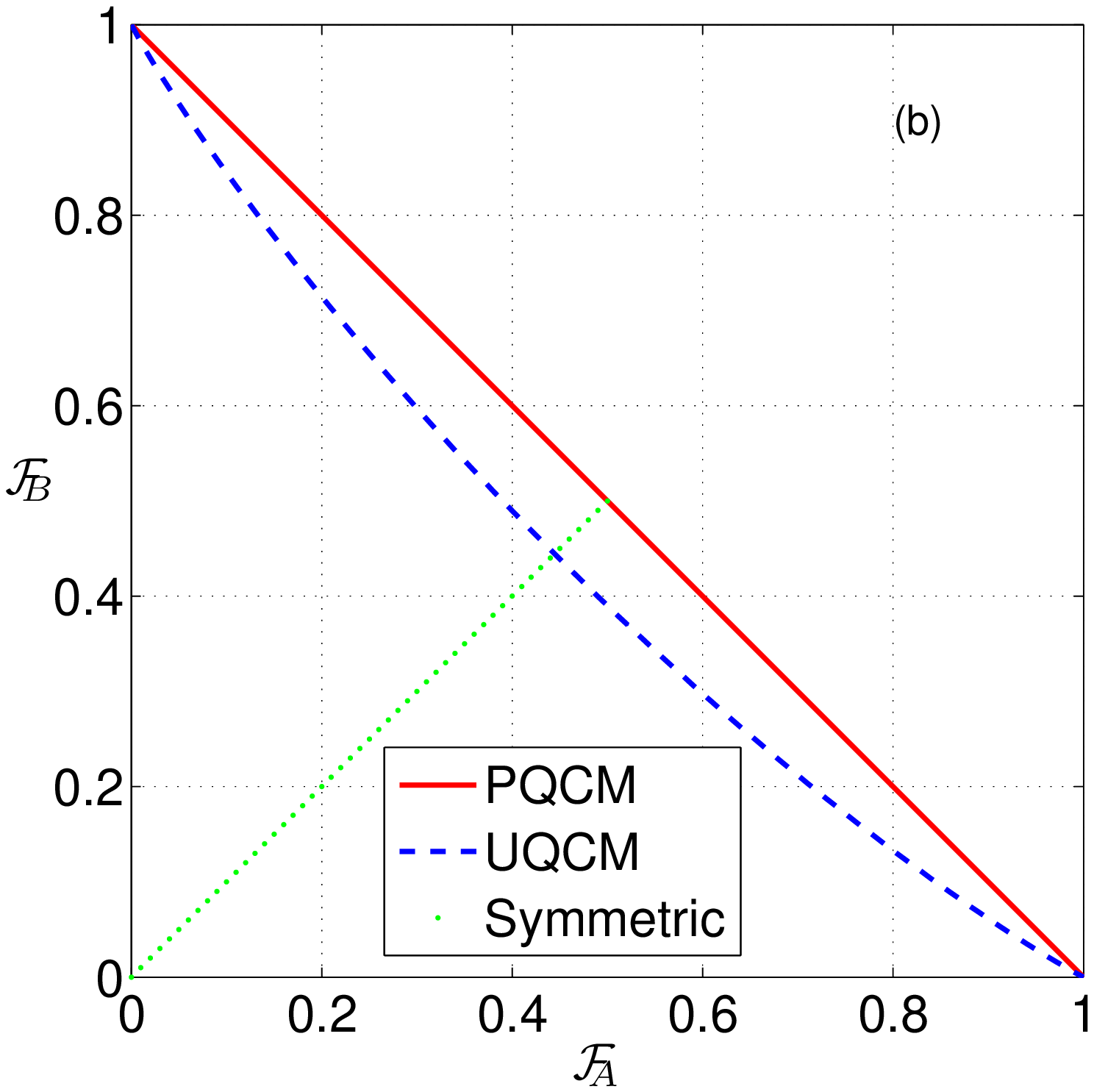}\\
  \caption{Trade-off relations for 2-dimensional asymmetric UQCM and PQCM: (a) Fidelity (b) QFI. The dotted lines denote
  the symmetric cases.
}\label{fig2}
\end{figure}

Previous studies suggest that the equatorial qubit state PQCM can be
realized by both economic\ucite{niu99} and
non-economic\ucite{bruss00,griff97} transformations. One can check that both economic and non-economic
methods achieve the same distributability of QFI. In the text we
discuss the non-economic case as it can be directly generalized to
\emph{d}-dimensional PQCM.  The transformation
of asymmetric PQCM can be written in the following form
\begin{align}
\label{e15}
&|i\rangle_{A}|0\rangle_{B}|\Sigma\rangle_{R}\rightarrow c|i\rangle_{A}|i\rangle_{B}|\Sigma_{i}\rangle_{R}\nonumber\\
&+\left(a|i\rangle_{A}|j\rangle_{B}+b|j\rangle_{A}|i\rangle_{B}\right)|\Sigma_{j}\rangle_{R},
\end{align}
with $i,j=0,1$ and $i\neq j$. $a^2+b^2+c^2=1$ is the normalization condition.
$|\Sigma_{i}\rangle_{R}$ is a set of orthogonal normalized ancillary state.
For the equatorial
qubit-state $|\psi\rangle=(|0\rangle+e^{i\theta}|1\rangle)/\sqrt{2}$, the output states have the form as
\begin{align}
\label{e16}
\rho_{A}&=2ac|\psi\rangle\langle\psi|+\frac{1-2ac}{2}\textrm{I}_{2},\\
\label{e17}
\rho_{B}&=2bc|\psi\rangle\langle\psi|+\frac{1-2bc}{2}\textrm{I}_{2}.
\end{align}
Then, the shrinking factors are $\eta_{A}=2ac, \eta_{B}=2bc$. Using the normalization condition, the shrinking factors
can be simplified as
\begin{align}
\label{e18}
&\eta_{A}=2a\sqrt{1-a^2-b^2},\\
\label{e19}
&\eta_{B}=2b\sqrt{1-a^2-b^2}.
\end{align}
As is seen, in the scenario of asymmetric PQCM, there are two free parameters to be optimized. Therefore,
an optimal asymmetric PQCM is defined as the following: if we fix the quality of one copy, then the other copy is optimal
with  the highest quality. From the Eqs. (\ref{e18}) and (\ref{e19}), one can eliminate $b$ and obtain the trade-off relation
between $\eta_{A}$ and $\eta_{B}$
\begin{align}
\label{e20}
\eta_{B}=\frac{\eta_{A}}{a}\sqrt{1-\frac{\eta_{A}^2}{4a^2}-a^2}.
\end{align}
Assuming $\eta_{A}$ is constant, the optimal value of $a_{optimal}=\eta_{A}/\sqrt{2}$
can be found, and the optimal trade-off relation reduces to
\begin{equation}
\label{e21}
\eta_{A}^2+\eta_{B}^2=1.
\end{equation}
The corresponding QFIs are
\begin{equation}
\mathcal{F}_{A}=\eta_{A}^2, \mathcal{F}_{B}=\eta_{B}^2
\end{equation}
In the symmetric case, we have $\eta_{A}=\eta_{B}=1/\sqrt{2}$. Remarkably, according to Eq. (\ref{e11}),
one can immediately find the inequality
\begin{equation}
\label{e22}
\mathcal{F}^{PQCM}=\mathcal{F}_{A}+\mathcal{F}_{B}=1\geq\mathcal{F}^{UQCM}.
\end{equation}
The meanings of above equation are twofold. On one hand,
it indicates that asymmetric PQCM performs better than asymmetric UQCM in
distributing QFI by virtue of the known information. This result is essentially in agreement with that of fidelity.
To be clear, we plot the trade-off relations between A's fidelity (QFI) and B's
fidelity (QFI) for both asymmetric UQCM and PQCM in Fig.~\ref{fig2}.
It is evident that the lines of asymmetric PQCM are always above those of asymmetric UQCM, except for the
start points and end points.
On the other hand, it should be noted that, for the 2-dimensional PQCM, the asymmetric case is as
good as the symmetric case on the capability of distributing QFI,
while the later always performs better than the former with the
measure of fidelity. The reason is that the fidelity $F_{B}$ as a function of $F_{A}$ is strictly
concave, but the QFI $\mathcal{F}_{B}$ as a function of $\mathcal{F}_{A}$ is convex. This results in
the sum of two fidelities and QFIs achieving its maximal and minimal value, respectively, in the symmetric case.

%In the symmetric case, we have $\eta_{A}=\eta_{B}=1/\sqrt{2}$.
%Comparing with the asymmetric UQCM whose optimal trade-off is the
%part of an ellipse in the quadrant I,  while the trade-off of
%asymmetric PQCM is the part of a circle in the $(\eta_{A},\eta_{B})$
%plane, as shown in Fig.~\ref{fig2}a. Thus, it is expected that the
%asymmetric PQCM would perform better than asymmetric UQCM in
%distributing QFI. That is true since one can immediately find
%\begin{equation}
%\label{e22}
%\mathcal{F}^{PQCM}=\mathcal{F}_{A}+\mathcal{F}_{B}=1\geq\mathcal{F}^{UQCM}.
%\end{equation}
%As seen from the $(\mathcal{F}_{A},\mathcal{F}_{B})$ plane, the
%trade-off relation of asymmetric PQCM is a line, while that of
%asymmetric UQCM is an arc below it, as shown in Fig.~\ref{fig2}a.
%Moreover, it should be emphasized that the asymmetric PQCM is as
%good as symmetric PQCM on the capability of distributing QFI.
%However, the later always performs better than the former with the
%measure of fidelity, as seen from Fig.~\ref{fig2}b.

\textbf{QFI distribution for \emph{d}-dimensional cloning.} Until now, we have
restricted our discussions to the 2-dimensional cloning. Although all
quantum information tasks can be performed by using only two-level systems, it
has been recently recognized that higher-dimensional quantum states
(i.e., qudits) can offer significant advantages for improving the
security of quantum cryptographic protocols\ucite{nikol05},
achieving higher information-density coding\ucite{durt04,ding13} and
reducing the required resources for quantum computation and
simulation\ucite{lanyon09,neeley09}.

Based on these considerations, it would be essential to extend
above discussions to \emph{d}-dimensional cloning. One may think the results will be trivial
and analogous conclusions will be obtained as well as the 2-dimensional cloning. However,
as we will show below, there are some similarities between them, but more importantly,
significant differences will appear with increasing dimensionality $d$.

\textbf{Asymmetric $d$-dimensional UQCM.} The optimal asymmetric
$d$-dimensional UQCM was proposed by Cerf\ucite{cerf00} and
Braunstein \emph{et al}\ucite{braun01}. For a $d$-dimensional
quantum system, the corresponding asymmetric UQCM can be generalized
directly from the transformation (\ref{e6}) with $|\Phi^{+}\rangle$
and $|+\rangle_{R}$ instead defined in higher-dimension,
$|\Phi^{+}\rangle=\frac{1}{\sqrt{d}}\sum_{j=0}^{d-1}|jj\rangle$, and
$|+\rangle_{R}=\frac{1}{\sqrt{d}}\sum_{j=0}^{d-1}|j\rangle_{R}$
respectively. Hence, the normalization condition now reads
$a^2+b^2+2ab/d=1$. The output reduced density matrices are written
in the form of Eq. (\ref{e3})
\begin{align}
\label{e23}
\rho_{A}&=(1-b^2)|\psi\rangle\langle\psi|+\frac{b^2}{d}\textrm{I}_{d},\\
\label{e24}
\rho_{B}&=(1-a^2)|\psi\rangle\langle\psi|+\frac{a^2}{d}\textrm{I}_{d},
\end{align}
with shrinking factors $\eta_{A}=1-b^2$ and $\eta_{B}=1-a^2$. In particular, if $a^2=b^2=d/(2d+2)$, we recover the results of symmetric UQCM. Similarly,
we can obtain a trade-off relation between two shrinking factors $\eta_{A}$ and $\eta_{B}$.
\begin{equation}
\label{e25}
\eta_{A}^2+\eta_{B}^2+\frac{2d^2-4}{d^2}(1-\eta_{A})(1-\eta_{B})\leq1,
\end{equation}
which corresponds to a set of ellipses in the space of shrinking factors that their eccentricities vary with dimensionality.
It should be noted that in the infinite dimensional case, the corresponding ellipse shrinks to the line $\eta_{A}+\eta_{B}=1$.

Now we turn to the calculation of QFI. Assuming the input state is a $d$-dimensional equatorial state,
then the QFIs of output states (\ref{e23}) and (\ref{e24}) are obtained directly by (\ref{e5}).
\begin{align}
\label{e26}
&\mathcal {F}_{A}=\frac{4(d-1)(1-b^2)^2}{2d+d(d-2)(1-b^2)},\\
\label{e27}
&\mathcal {F}_{B}=\frac{4(d-1)(1-a^2)^2}{2d+d(d-2)(1-a^2)}.
\end{align}
The tradeoff relation between $\mathcal {F}_{A}$ and $\mathcal {F}_{B}$ can be derived by substituting
\begin{equation}
\label{e28}
\eta_{X}=\frac{(d^2-2d)\mathcal {F}_{X}+\sqrt{(d^2-2d)^2\mathcal {F}_{X}^2+32(d^2-d)\mathcal {F}_{X}}}{8(d-1)}
\end{equation}
into (\ref{e25}) with $X=A,B$, but it is too complicated to present in the text.
Nevertheless, there is no doubt that one cannot gain, at the same time, two copies
whose QFIs are above values allowed by the trade-off relation.

\begin{figure}

  \centering
  \includegraphics[width=0.4\textwidth]{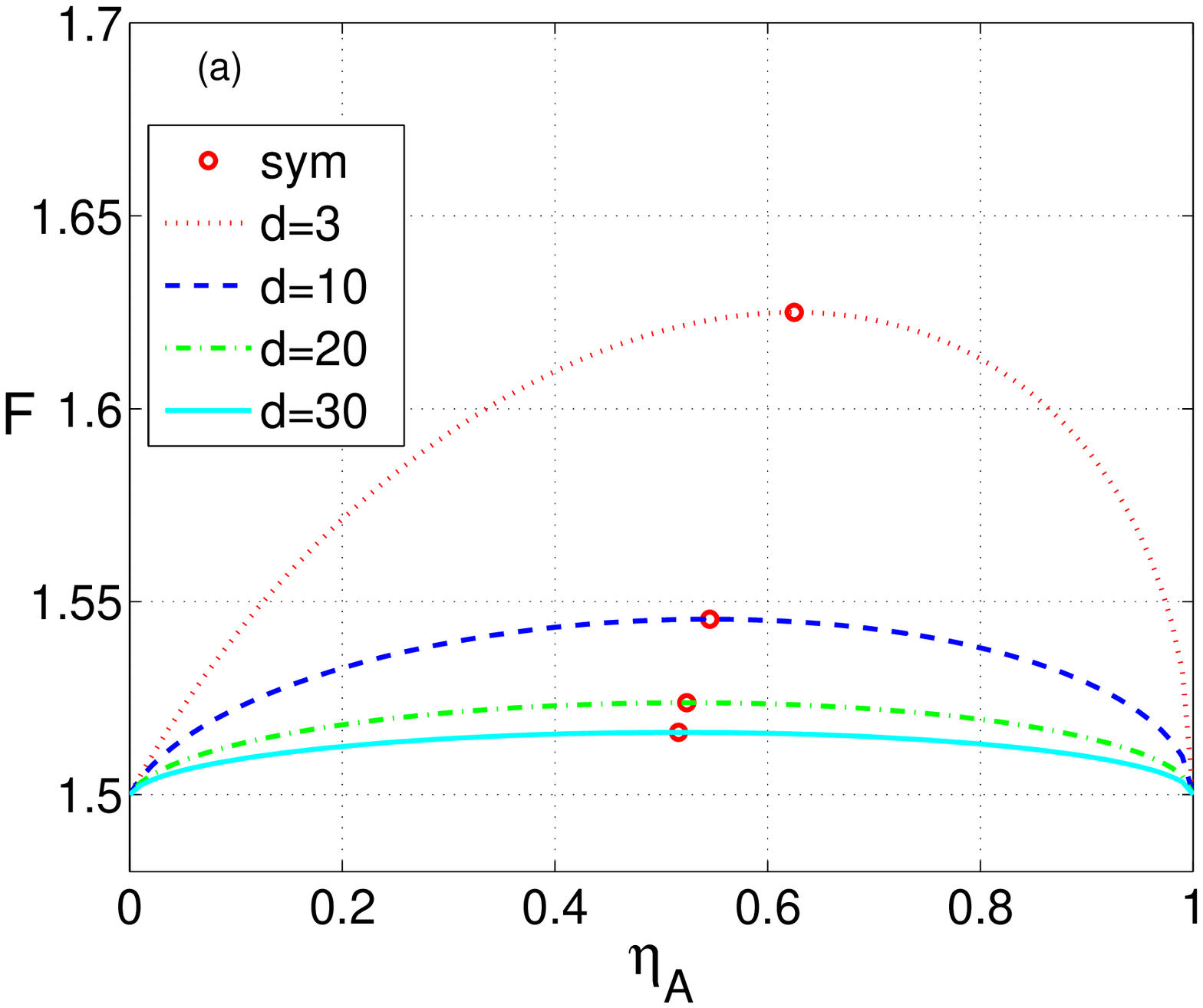}\\
  \includegraphics[width=0.4\textwidth]{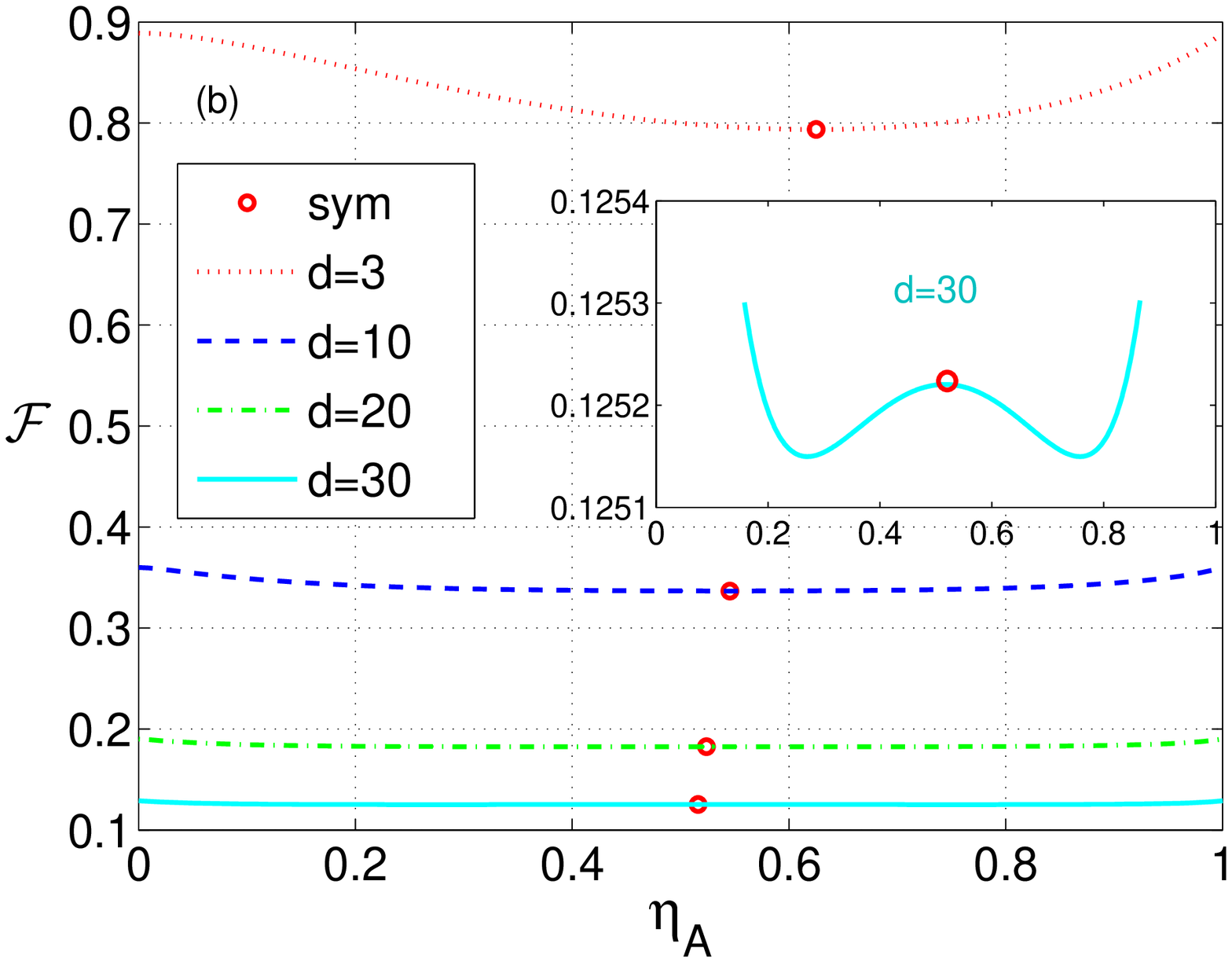}\\
  \caption{(a) Fidelity and (b) QFI as a function of $\eta_{A}$ for $d=3, 10, 20, 30$, respectively from top to down.
  The red circles denote the corresponding results of symmetric UQCM, and the insertion is magnified plot of $d=30$.}\label{fig3}
\end{figure}

We are concerned with whether the $d$-dimensional asymmetric UQCM
still performs better than symmetric UQCM in distributing QFI.
Against all expectations, the results become subtle with increasing
$d$. Unlike the 2-dimensional case where asymmetric UQCM is always better
than symmetric UQCM, the $d$-dimensional asymmetric UQCM may be
worse than symmetric UQCM under certain conditions. As shown in
Fig.~\ref{fig3}a, we find the sum of two fidelities still reaches its
largest value at the point of $\eta_{A}=\eta_{B}=(d+2)/(2d+2)$,
which means, by the measure of fidelity, the symmetric UQCM will
optimally copy the state regardless of the dimension\ucite{ghiu03}.
However, Fig.~\ref{fig3}b shows that the distributability of QFI
achieves its smallest value at the point of $\eta_{A}=\eta_{B}$ when
$d=3,10$, while it is interesting to note that the point of
$\eta_{A}=\eta_{B}$ becomes a local maximum point when $d=30$.
Namely, the asymmetric UQCM is no longer always better than
symmetric UQCM with increasing $d$. The reason is that even though
QFI is a monotonically function of the shrinking factor, it is not a
linear function of it. This sophisticated relation between
$\mathcal{F}$ and $\eta$ reveals above interesting results.

\begin{figure}

  \centering
  \includegraphics[width=0.4\textwidth]{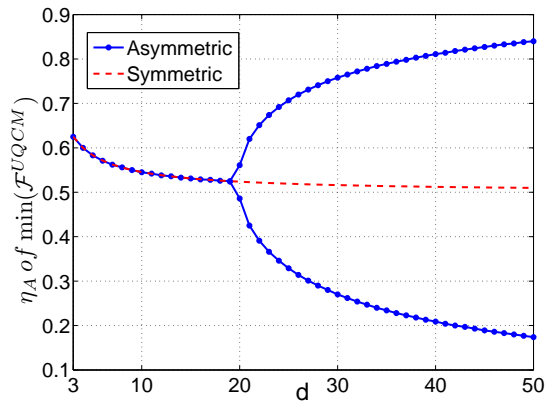}\\
  \caption{ $\eta_{A}$ of minimal $\mathcal{F}^{UQCM}$ as a function of dimensionality $d$.}\label{fig4}
\end{figure}

Naturally, we start wondering when the asymmetric UQCM may become
worse than symmetric UQCM. The numerical simulation shows that when
$d\leq18$, the $\eta_{A}$ of global minimal $\mathcal{F}$ is equal to the
symmetric case. While a bifurcation appears at $d=18$, which means
$\eta_{A}=(d+2)/(2d+2)$ is no longer the global point of minimal
$\mathcal{F}$,  as shown in Fig.~\ref{fig4}. Mathematically, we can
understand the bifurcation as follows: $\mathcal{F}$ as a function
of $\eta_{A}$ has three extreme points which are physically allowed
when $d\leq18$, and $\eta_{A}=\eta_{B}$ is the point of global minima.
When $d>18$, it has five extreme points, as seen from the insertion in Fig.~\ref{fig3}b.
Moreover, $\eta_{A}=\eta_{B}$ is no longer the global
minima but a local maximum point. Thus, the symmetric UQCM may outperform the asymmetric case.
However, it is hard to understand why the critical point is $d=18$ in physical, we
conjecture this critical point is related to the Hilbert space structure of qudits.
We leave this as an open question and the further study is underway.

%specified as the following unitary transformation
%$a+b=c$. If a $d$-dimensional equatorial state of Eq. (\ref{e4}) is given to be copied, then the output reduced
%states $\rho_{A}$ and $\rho_{B}$ have the form of Eq. (\ref{e5}) with the shrinking factors
%\begin{align}
%\eta_{A}=(d-2)a^2+2ac,\\
%\eta_{B}=(d-2)b^2+2bc.
%\end{align}

\textbf{Asymmetric $d$-dimensional PQCM.}
The generalization of asymmetric PQCM to $d$-dimensional is much more difficult, and in particular, it is too complicated to present
an analytical trade-off relation between two copies. However, with the aid of numerical simulations, we can confirm
two main results about the distribution of QFI in asymmetric $d$-dimensional PQCM:
(i) PQCM gains an advantage over UQCM by utilizing the priori information, and
(ii) a sudden change of the point of minimum also exists in asymmetric PQCM with increasing $d$.

\begin{figure}

  \centering
  \includegraphics[width=0.4\textwidth]{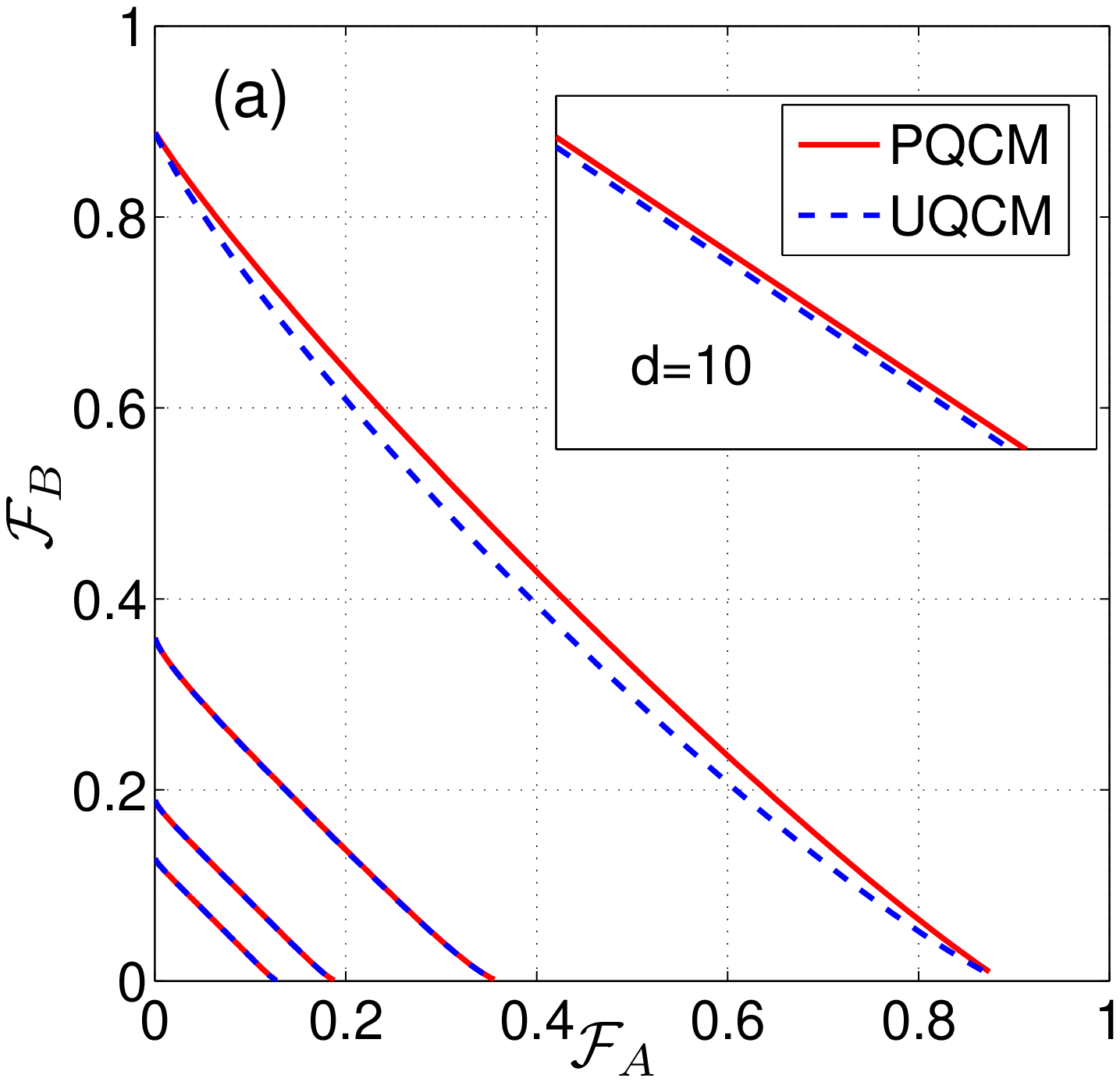}\\
   \includegraphics[width=0.4\textwidth]{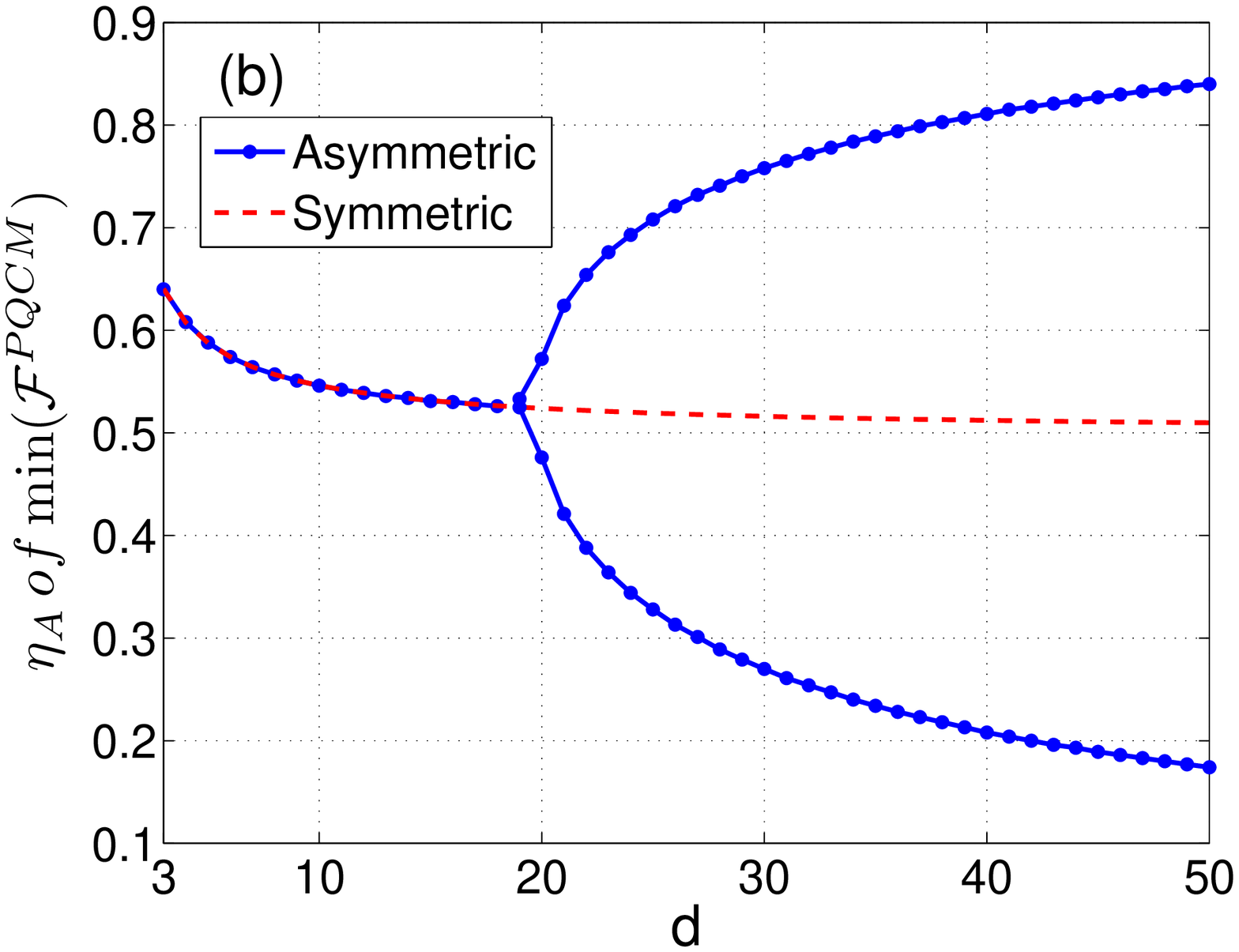}\\
  \caption{(a) Trade-off relations between $\mathcal{F}_{A}$ and $\mathcal{F}_{B}$
  for \emph{d}-dimensional asymmetric UQCM (dashed lines) and PQCM (solid lines)
  with $d=3,10,20,30$, respectively from top to bottom. The insertion is magnified
  plot of $d=10$. Note that when $d=20$ and 30, the two lines overlap greatly.
  (b) $\eta_{A}$ of minimal $\mathcal{F}^{PQCM}$ as a function of dimensionality $d$.
}\label{fig5}
\end{figure}

The cloning transformation of asymmetric $d$-dimensional PQCM  can
be introduced\ucite{reza05}
\begin{align}
\label{e29}
&|i\rangle_{A}|0\rangle_{B}|\Sigma\rangle_{R}\rightarrow c|i\rangle_{A}|i\rangle_{B}|\Sigma_{i}\rangle_{R}\nonumber\\
&+\left(a\sum_{i\neq j}^{d-1}|i\rangle_{A}|j\rangle_{B}+b\sum_{i\neq j}^{d-1}|j\rangle_{A}|i\rangle_{B}\right)|\Sigma_{j}\rangle_{R},
\end{align}
with the normalization condition $(d-1)(a^2+b^2)+c^2=1$. Given the input state in the form of (\ref{e4}), then, the shrinking factors of
the output copies read as
\begin{align}
\label{e30}
&\eta_{A}=(d-2)a^2+2a\sqrt{1-(d-1)(a^2+b^2)},\\
\label{e31}
&\eta_{B}=(d-2)b^2+2b\sqrt{1-(d-1)(a^2+b^2)},
\end{align}
where we have use the normalization condition to eliminate the parameter $c$. Similar to the 2-dimensional case,
here we again need to optimize two free tuning parameters. Therefore, in the same way, an optimal asymmetric PQCM is defined by optimizing
$\eta_{B}$ as large as possible when $\eta_{A}$ is fixed, and vice versa. By eliminating the parameter $b$, we can obtain
the trade-off relation (not the optimal one)
\begin{align}
\label{e32}
\eta_{B}=&\frac{d-2}{d-1}\bigg[1-(d-1)a^2-\big(\frac{\eta_{A}-(d-2)a^2}{2a}\big)^2\bigg]\nonumber\\
&+\frac{\eta_{A}-(d-2)a^2}{a}\sqrt{\frac{1-\big(\frac{\eta_{A}-(d-2)a^2}{2a}\big)^2}{d-1}-a^2}.
\end{align}
The optimal trade-off relation need to be further optimized by choosing a proper value of $a_{optimal}$ to make the largest $\eta_{B}$.
Unfortunately, there is not a closed analytical form of $a_{optimal}$ for any dimensionality $d$. However, by simple numerical simulations,
we find that asymmetric PQCM indeed always performs better than asymmetric UQCM in distributing QFI as shown in Fig.~\ref{fig5}a. When the dimensionality \emph{d} is large
(e.g., $d=20$), it should be noted that the advantage of PQCM
over UQCM almost disappears. Moreover, similar to
the case of asymmetric $d$-dimensional UQCM, the asymmetric $d$-dimensional PQCM is not always better than the symmetric case for any dimensionality $d$.
Fig.~\ref{fig5}b shows that a bifurcation of the point of global minimum also occurs at $d=18$.
This phenomena stresses that the critical point appearing at $d=18$ is not in any sense accidental. The physical reason
behind this is worth further study.

\textbf{Discussions.}
In this paper, we have investigated the distribution of QFI in asymmetric cloning machines which produce two nonidentical copies.
In particular, we have elucidated four questions as we mentioned before.
Here, we summarize our results by replying these questions.
(i) The answer is YES. It is definite that improving the QFI in one copy results in decreasing the QFI of
the other copy, and the trade-off relation can be obtained analytically except for the asymmetric $d$-dimensional PQCM.
(ii) The answer is also YES. Thanks to a priori knowledge of the input states, PQCM always
performs better than UQCM in distributing QFI. (iii) The answer is not so straightforward. It should be divided into two
categories: for 2-dimensional cloning, asymmetric cloning always outperforms symmetric cloning on the distribution of QFI;
While for the \emph{d}-dimensional cloning case, the above conclusion only holds when $d\leq18$ and becomes invalid when $d>18$, i.e., the asymmetric
cloning is not always better than symmetric cloning for any dimensionality.
(iv) The most significant difference between fidelity and QFI is that fidelity is a linear function of the shrinking
factor while QFI is nonlinear. This leads to the counterintuitive result that symmetric cloning is always optimal from the perspective
of fidelity, but asymmetric cloning usually works better than symmetric cloning
on the distribution of QFI, except for some particular situations (e.g., when $d=30$ and $\eta_{A}=0.25$, see the insertion in Fig.~\ref{fig3}b).

In view of these findings, we note that there are some problems in need of further clarifications. The first important issue is to understand why does the critical point appear
at $d=18$, not other numbers. Secondly, we should realize that we have confined our discussion to
the distributability of single parameter in asymmetric cloning machines. However,
from both theoretical and practical points of view, it seems to
be interesting to examine the problem of multi-parameter distribution in asymmetric quantum cloning.
Intuitively, there would be a trade-off relation of the quantum Fisher information matrices between two nonidentical copies.
These would be very intriguing topics that need
further studies.

%we should argue that we have confined our discussion to
%the distributability of asymmetric cloning machines. Thus, it is reasonable to choose $\mathcal{F}=\mathcal{F}_{A}+\mathcal{F}_{B}$
%as the measure of distributability. However, if one try to estimating the parameter as precise as possible from two nonidentical
%copies, an effective QFI $\mathcal{F}_{eff}=\mathcal{F}_{A}\mathcal{F}_{B}/(\mathcal{F}_{A}+\mathcal{F}_{B})$ based on the quantum Carmel-Rao bound
%would be a meaningful measure.

\section*{\textcolor{black}{Methods\label{sec:Methods}}}

Here, we give the details of the derivation of Eq.(\ref{e5}) from Eqs. (\ref{e3}) and (\ref{e4}). To be clear,
we can rewrite the (\ref{e3}) as
\begin{equation}
\rho=\frac{(d-1)\eta+1}{d}|\psi\rangle\langle\psi|+\frac{1-\eta}{d}(\textrm{I}_{d}-|\psi\rangle\langle\psi|).
\end{equation}
Note that the eigenvalues of $\rho$ consists of only two categories: $\lambda_0=[(d-1)\eta+1]/d$,
and $\lambda_{n}=(1-\eta)/d$ with $1\leq n\leq d-1$. Obviously, $|\psi_0\rangle=|\psi\rangle$ is an eigenstate of $\rho$. Thus the problem is converted to construct
a complete orthogonal set (containing $d-1$ bases) of the operator $\hat{\Pi}=\texttt{I}-|\psi\rangle\langle\psi|$ which is also orthogonal
to $|\psi\rangle$ at the same time. The procedure can be divided into three steps: (i) finding $d-1$ bases of $\hat{\Pi}$ which are orthogonal to $|\psi\rangle$;
(ii) using the Gram-Schmidt procedure to orthogonalize them and (iii) the normalization.

\emph{Step} (i) Intuitively, the $d-1$ bases which are orthogonal to $|\psi\rangle$ can be written as
\begin{equation}
|\varphi_{n}\rangle=\frac{1}{\sqrt{2}}\big[-e^{-i(\theta_{n}-\theta_0)},0\ldots,\underbrace{1}_{\textrm{nth}},\ldots0\big]
\end{equation}
where $1\leq n\leq d-1$.

\emph{Step} (ii) According to the procedure of Gram-Schmidt orthonormalization, one can
construct a set of orthogonal but un-normalized bases:
\begin{equation}
|\widetilde{\psi}_n\rangle=|\varphi_n\rangle-\frac{1}{n}\sum_{m=1}^{n-1}e^{i(\theta_{m}-\theta_n)}|\varphi_j\rangle.
\end{equation}

\emph{Step} (iii) Notice that $\langle\widetilde{\psi}_{n}|\widetilde{\psi}_n\rangle=(n+1)/2n$, we finally obtain the
$d-1$ orthogonal and normalized bases of the operator $\hat{\Pi}$
\begin{equation}
|\psi_{n}\rangle=\sqrt{\frac{2n}{n+1}}\bigg[|\varphi_n\rangle-\frac{1}{n}\sum_{m=1}^{n-1}e^{i(\theta_{m}-\theta_n)}|\varphi_j\rangle\bigg],
\end{equation}
which are also orthogonal to $|\psi\rangle$.

By this time, we have diagonalized the state (\ref{e3}) in the bases $\{|\psi\rangle,|\psi_{1}\rangle,\ldots,|\psi_{n}\rangle\}$,
and then the QFI can be calculated by Eq. (\ref{e2}). Note that all the parameters $\theta_{k}$ is equally weighted due to the symmetry
of $|\psi\rangle$, thus the QFI of any parameter is the same. Furthermore, we observe that the part of classical Fisher information vanishes since
the probability distribution is independent of parameters $\theta_{k}$, if measured in this bases. The remaining work is to determine the last two terms in Eq. (\ref{e2}).
After lots of complicated but straightforward calculations, we obtain the QFI of any parameter $\theta_{k}$
\begin{align}
\mathcal {F}_{\theta_k}&=\sum_{n}\lambda_{n}\mathcal{F}_{\theta,n}
-\sum_{n\neq m}\frac{8\lambda_{n}\lambda_{m}}{\lambda_n+\lambda_m}|\langle\psi_{n}|\partial_{\theta}\psi_{m}\rangle|^2,\nonumber\\
&=\frac{4(d-1)\eta^2}{2d+d(d-2)\eta}.
\end{align}

\acknowledgments
The authors are supported by the National Natural Science Foundation of China under
Grants No. 11247006, 11025527 and 10935010, the National 973 program under Grants No. 2012CB921602,
and the China Postdoctoral Science Foundation under Grant No. 2014M550598.

%\textbf{Author Contributions}
%
%X.X. and Y.Y. proposed the idea and carried out the calculations under the guidance of X.G.W.
%L.M.Z made some numerical simulations and plotted the figures. X.X wrote the paper. All authors discussed the results and commented on the manuscript.
%
%
%
%\textbf{Additional information}
%
%The authors declare no competing financial interests.

%%%%%%%%%%%%%%%%%%%%%%%%%%%%%%%%%%%%%%%%%%%%%%%%%%%%%%%%%%%%%%%%%%%%%%%%%%%%%%%%%%%%%%%%%%%%%

\end{document}